\documentclass[a4paper,11pt,times,authoryears,reqno]{amsart}

\usepackage{geometry}
\usepackage{amsmath}
\usepackage{bm}
\usepackage{amssymb}
\usepackage{graphicx}
\usepackage{mathrsfs}
\usepackage{subfig}
\usepackage{float}
\usepackage{enumerate}
\usepackage{booktabs}
\usepackage{adjustbox}
\usepackage[colorlinks,citecolor=blue]{hyperref}
\usepackage[square,sort,comma,numbers]{natbib}

\newtheorem{theorem}{Theorem}[section]
\newtheorem{lemma}[theorem]{Lemma}

\theoremstyle{definition}

\theoremstyle{remark}

\numberwithin{equation}{section}

\def\bH{{\boldsymbol{H}}} 
\def\bh{{\boldsymbol{h}}}

\def\bA{{\mathbf A}}

\def\bC{{\mathbf C}}
\def\bD{{\mathbf D}}

\def\bF{{\mathbf F}}
\def\bI{{\mathbf I}} 
\def\bM{{\mathbf M}}
\def\bN{{\mathbf N}}
\def\bQ{{\mathbf Q}}
\def\bW{{\mathbf W}}
\def\hW{\widehat{\bW}}

\def\bX{{\mathbf X}}

\def\bb{{\mathbf b}}  
\def\bx{{\mathbf x}} 
\def\bu{{\mathbf u}} 

\def\bv{{\mathbf v}} 
\def\by{{\mathbf y}} 

\def\bz{{\mathbf z}} 
\def\hz{\widehat{\bz}}

\def\by{{\mathbf y}}

\def\balpha{{\boldsymbol{\alpha}}} 
\def\bpi{{\boldsymbol{\pi}}} 
\def\hbpi{\widehat{\boldsymbol \pi}}

\def\bPsi{{\boldsymbol{\Psi}}}
\def\bLambda{{\boldsymbol{\Lambda}}}

\def\bDelta{{\boldsymbol{\Delta}}}

\def\tbeta{\widetilde{\bbeta}}
\def\bbeta{{\boldsymbol{\beta}}}
\def\hbbeta{\widehat{\boldsymbol \beta}}
\def\hbz{\widehat{\bz}}

\def\no{\nonumber}

\def\bzero{{\mathbf 0}}

\def\mle{\bm{\widehat\beta}_{\textrm{MLE}}}

\def\lle{\bm{\widehat\beta}_{\textrm{LE}}}
\def\lte{\bm{\widehat\beta}_{\textrm{LTE}}}
\def\sre{\bm{\widehat\beta}_{\textrm{SRE}}}
\def\srle{\bm{\widehat\beta}_{\textrm{SRLE}}}
\def\srlte{\bm{\widehat\beta}_{\textrm{SRLTE}}}

\title{On the stochastic restricted Liu-type maximum likelihood estimator in logistic regression model}
\author{Jibo Wu*}
\author{Yasin Asar}

\address[Jibo Wu]{Corresponding Author: Key Laboratory of Group \& Graph Theories and Applications, Chongqing University of Arts and Sciences, Chongqing, 402160, China}
\email{linfen52@126.com}

\address[Yasin Asar]{ Department of Mathematics-computer Sciences, Necmettin Erbakan University, Konya, Turkey }
\email{yasar@konya.edu.tr, yasinasar@hotmail.com }

\date{}

\begin{document}

\maketitle

\begin{abstract} In order to overcome multicollinearity, we propose a stochastic restricted Liu-type maximum likelihood estimator by incorporating Liu-type maximum likelihood estimator (Inan and Erdogan, 2013) to the logistic regression model when the linear restrictions are stochastic. We also discuss the properties of the new estimator. Moreover, we give a method to choose the biasing parameter in the new estimator. Finally, a simulation study is given to show the performance of the new estimator. 
\end{abstract}

\noindent 

\section{Introduction}

\noindent Consider the following logistic regression model
\begin{equation} \label{log_model} 
y_{i} =\pi _{i} +\varepsilon _{i} ,i=1,\ldots ,n, 
\end{equation} 
where 
$\pi_{i} =\pi \left(x_{i} \right)=E\left[y_{i} \right]= \frac{e^{x_{i}} \bbeta}{1+e^{x_{i}} \bbeta } $

$y_{i} \sim Bernoulli\left(\pi _{i} \right)$ 
and $\bbeta =\left(\beta _{0} ,\beta _{1} ,\ldots ,\beta _{p} \right)^{\top} $ denotes the unknown (p+1)-vector of regression coefficients, $\bX=\left(\bx_{1} ,\ldots ,\bx_{n} \right)^{\top} $ is the $n\times \left(p+1\right)$ data matrix with $\bx_{i} =\left(1,x_{1i} ,\ldots ,x_{pi} \right)^{\top} $ and $\varepsilon _{i} $'s are independent with zero mean and their variance equal to $w_{i} =\pi _{i} \left(1-\pi _{i} \right)$. 

In the estimation process of the coefficient vector $\bbeta $, one often uses the maximum likelihood (ML) approach. One may use the iteratively re-weighted least squares algorithm (IRLS) as follows (\citet{SK2013}) :
\begin{equation} \label{MLE_iteration} 
\hbbeta^{t+1} =\hbbeta^{t} +\left(\bX^{\top} \widehat{\bW} \bX\right)^{-1} \bX^{\top} \widehat{\bW}^{t} \left(\by-\hbpi^{t} \right) 
\end{equation} 
where $\hbpi^{t} $ is the estimated values of $\hbpi$ using $\hbbeta_{t} $ and $\widehat{\bW}^{t} = {\rm diag}\left(\hat{\pi }_{i}^{t} \left(1-\hat{\pi }_{i}^{t} \right)\right)$ such that $\hat{\pi }_{i}^{t} $ is the ith element of $\hbpi^{t} $. After some algebra, Equation \eqref{MLE_iteration} can be written as follows:
\begin{equation} \label{1.4} 
\hbbeta_{ML} = \left(\bX^{\top} \widehat{\bW} \bX\right)^{-1} \bX^{\top} \widehat{\bW} \hbz 
\end{equation} 
where $\hbz^{\top} =\left(z_{1} \cdots z_{n} \right)$ with $\eta _{i} =x'_{i} \beta $ and $z_{i} =\eta _{i} +(y_{i} -\pi _{i} )(\partial \eta _{i} / \partial \pi _{i} )$.

As it is well known that when the multicollinearity exists in the explanatory variables, the MLE becomes unstable and its variance is inflated. Many methods have been proposed to overcome this problem. We refer to the following papers: \citet{Manson2012}, \citet{IE2013}, \citet{WuAsar2016}, \citet{AG2016}. In this article, we also suppose that $\bbeta $ is subjected to lie in the sub-space restriction $\bH\bbeta =\bh$, where $\bH$ is $q\times \left(p+1\right)$ known matrix and $\bh$ is a $q\times 1$ vector of pre-specified values. This problem was also studied in  \citet{KS2011}, \citet{SK2013}, \citet{WuAsar2017}, \citet{NaWi2015}

For the stochastic linear restrictions, \citet{NaWi2015}, \citet{VW2016a} \citet{VW2016b} discussed the logistic regression model. In this paper, we will discuss the logistic regression model with stochastic linear restrictions. By combining the Liu-type maximum likelihood estimator (\citet{IE2013}) and stochastic linear restrictions, we propose a new estimator called stochastic restricted Liu-type maximum likelihood estimator to overcome multicollinearity. We also discuss the properties of the new estimator and we show that the new estimator can not only overcome multcollineariy, but also has good properties than other exists estimators. 

The rest of the paper is organized as follows, in Section 2, the new estimator is proposed and the properties of the new estimator is discussed in Section 3. A method is given to choose the biasing parameters in Section 4 and we also conduct a simulation study to show the performance of the new estimator in Section 5, some conclusion remarks are given in Section 6.

\section{The new estimator}

\noindent For the model \eqref{log_model}, \citet{Manson2012} proposed a Liu maximum likelihood estimator (LE) to overcome the multicollinearity, which is defined as
\begin{equation} \label{liu} 
\lle (d)=\left(\bX^{\top}\hW\bX+\bI \right)^{-1} \left(\bX^{\top}\hW\hz+d \mle \right)=\bF_{d} \mle  
\end{equation} 
where $\bF_{d} =\left(\bX^{\top}\hW\bX+\bI \right)^{-1} \left(\bX^{\top}\hW \bX+d\bI \right),0<d<1$.

\citet{IE2013} proposed a Liu-type maximum likelihood estimator (LTE), which is defined as
\begin{equation} \label{lte} 
\lte(k,d)=\left(\bX^{\top}\hW\bX+k\bI \right)^{-1} \left(\bX^{\top}\hW\hz-d \mle  \right)=\bF_{kd} \mle
\end{equation} 
where $\bF_{kd} =\left(\bX^{\top}\hW\bX+k\bI \right)^{-1} \left(\bX^{\top}\hW\bX-d\bI\right),k>0,-\infty <d<+\infty $.

Assume that the following general stochastic linear restrictions is given in addition to the general logistic regression model
\begin{equation} \label{rest} 
\bh=\bH\bbeta +\bv, E\left(\bv \right)=\bzero ,Cov\left(\bv \right)= \bPsi  
\end{equation} 
where $\bH$ is $q\times \left(p+1\right)$ known matrix and $\bh$ is a $q\times 1$ vector of pre-specified values. $\bv$ is an random vector. We assumed that $\bv$ is independent of $\boldsymbol{\varepsilon} $.

By combining \eqref{log_model} and \eqref{rest}, \citet{NaWi2015} proposed the following stochastic restricted maximum likelihood estimator (SRE) as follows:
\begin{equation} \label{sre} 
\sre =\left(\bC+\bH^{\top}\bPsi ^{-1} \bH\right)^{-1} \left(\bX^{\top}\hW\hz+\bH^{\top}\bPsi ^{-1} \bh\right) .
\end{equation} 
    \citet{VW2016b} by combining the LE and SRE, proposed the stochastic restricted Liu maximum likelihood estimator (SRLE)
\begin{equation} \label{srle} 
\srle \left(d\right)=\bF_{d} \left(\bC+\bH^{\top}\bPsi ^{-1} \bH\right)^{-1} \left(\bX^{\top}\hW\hz+\bH^{\top}\bPsi  ^{-1} \bh\right)=\bF_{d} \sre . 
\end{equation} 
    Now incorporating the LTE to the logistic regression model under the stochastic linear restriction, we propose a new biased estimator which is called stochastic restricted Liu-type maximum likelihood estimator (SRLTE)
\begin{equation} \label{srlte} 
\srlte \left(k,d\right)=\bF_{kd} \left(\bX^{\top}\hW\bX+\bH^{\top}\bPsi ^{-1} \bH\right)^{-1} \left(\bX^{\top}\hW\hz+\bH^{\top}\bPsi \bh\right)=\bF_{kd} \sre 
\end{equation} 
where $\bF_{kd} =\left(\bX^{\top}\hW\bX+k\bI\right)^{-1} \left(\bX^{\top}\hW\bX-d\bI\right)$, $k>0,-\infty <d<+\infty $ are two biasing parameters. By the definition of SRLTE, we can see that it is a general estimator which includes the MLE, LE, LTE, SRE and SRLE as special cases.

\begin{itemize}
\item When $k=-d$, $\srlte \left(k,d\right)=\sre =\left(\bX^{\top}\hW\bX+\bH^{\top}\bPsi ^{-1} \bH\right)^{-1} \left(\bX^{\top}\hW\hz+\bH^{\top}\bPsi ^{-1} \bh\right) $,

\item When $k=1,d=-d$,
$\srlte \left(k,d\right)=\srle =\bF_{d} \left(\bX^{\top}\hW\bX+\bH^{\top}\bPsi ^{-1} \bH\right)^{-1} \left(\bX^{\top}\hW\hz+\bH^{\top}\bPsi ^{-1} \bh\right)$,

\item When $\bH=\bzero$, $\srlte \left(k,d\right)=\lte=\bF_{kd} \mle$,

\item When $\bH=\bzero$, $k=-d$, $\srlte\left(k,d\right)=\mle $,

\item When $\bH=\bzero$, $k=1,d=-d$, $\srlte \left(k,d\right)=\lle =\bF_{d} \mle$.
\end{itemize}

In next section, we will study the performance of the new estimator over the existing estimator given in the literature.

\section{The properties of the new estimator}

\noindent In this section, we will present the comparison of the new estimator with other estimators under the mean squared error matrix criteria. Firstly, we present the definition of mean squared error matrix. The mean squared error matrix (MSEM) of estimator $\tbeta$ is defined as 
\begin{eqnarray*}
MSEM(\tbeta)=E[(\tbeta-\bbeta )(\tbeta-\bbeta )^{\top}=Cov(\tbeta)+Bias(\tbeta)Bias(\tbeta)^{\top}.
\end{eqnarray*}
where $Cov(\tbeta)$ is the covariance matrix and $Bias(\tbeta)$ is the bias of the estimator $\tbeta$.
The scalar mean squared error (SMSE) is defined as follows:
$$SMSE(\tbeta)=tr\left[MSEM(\tbeta)\right]$$
where tr is the trace of a matrix. 
Now, we will present three lemmas and use them to prove the theorems given in the next section.
\begin{lemma}\label{fare}
(\citet{Fare1976}, \citet{rao}) 
Suppose that $\bM$ be a positive definite matrix, namely $\bM>\bzero$, $\balpha$ be some vector, then $\bM-\balpha \balpha ^{\top} \ge \bzero$ if and only if $\balpha ^{\top} \bM^{-1} \balpha \le 1$.
\end{lemma}

\begin{lemma} \label{rao}
(\citet{rao}) 
Let $\bM>\bzero$, $\bN>\bzero$, then $\bM>\bN$, if and only if $\lambda _{\max } \left(\bN \bM^{-1} \right)<1$.
\end{lemma}

\begin{lemma} \label{lem3}
(\citet{rao}) 
Let $\hbbeta_{j} =\bA_{j} \by, j=1,2$ be two competing estimator of $\bbeta $. Assume that $\bDelta =Cov\left(\bbeta_{1} \right)-Cov\left(\hbbeta_{2} \right)>\bzero$, then $MSEM\left(\bbeta_{1} \right)-MSEM\left(\hbbeta_{2} \right)>\bzero$ if and only if $u^{\top}_{2} \left(\bDelta +\bu_{1} \bu^{\top}_{1} \right)^{-1} \bu_{2} \le 1$, where $\bu_{j} $ denotes the bias of $\hbbeta_{j} $.
\end{lemma}

\noindent 

\subsection{The new estimator versus SRE}

The asymptotic properties of the SRE are given as follows: The expected value and the covariance matrix are respectively given by
\begin{equation} \label{asymp_sre} 
E\left[\sre \right]=\bbeta , Cov\left[\sre \right]=\left(\bX^{\top}\hW \bX+\bH^{\top}\bPsi ^{-1} \bH\right)^{-1} . 
\end{equation} 
Then we may have the MSEM of SRE as
\begin{equation} \label{Gmmse_sre} 
MSEM\left[\sre \right]=\left(\bX^{\top}\hW \bX+\bH^{\top}\bPsi ^{-1} \bH\right)^{-1} . 
\end{equation} 
The asymptotic properties of the new estimator SRLTE are obtained similarly as follows:
\begin{equation} \label{e_srlte} 
E\left[\srlte \left(k,d\right)\right]=\bF_{kd} \beta =\left(\bX^{\top}\hW \bX+k\bI\right)^{-1} \left(\bX^{\top}\hW \bX-d\bI\right)\bbeta  
\end{equation} 
and
\begin{equation} \label{cov_srlte} 
Cov\left[\srlte  \left(k,d\right)\right]=\bF_{kd} \left(\bX^{\top}\hW \bX+\bH^{\top} \bPsi ^{-1} \bH \right)^{-1} \bF_{kd} .
\end{equation} 
Then MSEM of SRLTE is given by
\begin{equation} \label{mmse_srlte} 
MSEM\left[\srlte \left(k,d\right)\right]=\bF_{kd} \left(\bX^{\top}\hW \bX+\bH^{\top} \bPsi ^{-1} \bH\right)^{-1} \bF_{kd} +\bb_{1} \bb'_{1}  
\end{equation} 
where $\bb_{1} =(d+k)\left(\bX^{\top}\hW \bX+k\bI\right)^{-1} \bbeta $.

Now we consider the following difference
\begin{eqnarray} \label{d1} 
\bDelta _{1} &=& MSEM\left[\sre\right]-MSEM\left[\srlte \left(k,d\right)\right] \no \\ &=& \left(\bX^{\top}\hW \bX+\bH^{\top} \bPsi ^{-1} \bH \right)^{-1} -\bF_{kd} \left(\bX^{\top}\hW \bX+\bH^{\top} \bPsi ^{-1} \bH\right)^{-1} \bF_{kd} -\bb_{1} \bb^{\top}_{1}  \no \\ &=& \bD_{1} -\bb_{1} \bb^{\top}_{1} 
\end{eqnarray} 
where $\bD_1 = \left(\bX^{\top}\hW \bX+\bH^{\top} \bPsi ^{-1} \bH \right)^{-1} -\bF_{kd} \left(\bX^{\top}\hW \bX+\bH^{\top} \bPsi ^{-1} \bH\right)^{-1} \bF_{kd} $.

Since $\left(\bX^{\top}\hW \bX+\bH^{\top} \bPsi ^{-1} \bH \right)^{-1} >\bzero$, $\bF_{kd} \left(\bX^{\top}\hW \bX+\bH^{\top} \bPsi ^{-1} \bH\right)^{-1} \bF_{kd} >\bzero$, then by Lemma \ref{rao}, when $\lambda _{\max } \left\{F_{kd} \left(\bX^{\top}\hW \bX+\bH^{\top} \bPsi ^{-1} \bH\right)^{-1} F_{kd} \left(\bX^{\top}\hW \bX+\bH^{\top} \bPsi ^{-1} \bH\right)\right\}<1$, $\bD_{1} >\bzero$. Then by Lemma \ref{fare}, when $\bb^{\top}_{1} \bD_{1} ^{-1} \bb_{1} \le 1$, $\bDelta _{1} \ge \bzero$.

\noindent Based on the arguments above, we may have the following theorem:

\begin{theorem} When $\lambda _{\max } \left\{F_{kd} \left(\bX^{\top}\hW \bX+\bH^{\top} \bPsi ^{-1} \bH\right)^{-1} F_{kd} \left(\bX^{\top}\hW \bX+\bH^{\top} \bPsi ^{-1} \bH\right)\right\}<1$, the new estimator is superior to the SRE if and only if $\bb^{\top}_{1} \bD_{1} ^{-1} \bb_{1} \le 1$.
\end{theorem}

\noindent 

\subsection{The new estimator versus SRLE}

\noindent The asymptotic properties of the SRLE are obtained as follows:
\begin{equation} \label{srle} 
E\left[\srle \left(d\right)\right]=\bF_{d} \bbeta =\left(\bX^{\top}\hW \bX+\bI\right)^{-1} \left(\bX^{\top}\hW \bX+d\bI\right)\bbeta  
\end{equation} 
and
\begin{equation} \label{cov_srle} 
Cov\left[\hat{\bbeta }_{SRLE} \left(d\right)\right]=\bF_{d} \left(\bX^{\top}\hW \bX+\bH^{\top} \bPsi \bH \right)^{-1} \bF_{d}  
\end{equation} 
Then we have MSEM of SRLE as
\begin{equation} \label{mmse_srle} 
MSEM\left[\srle \left(d\right)\right]=\bF_{d} \left(\bX^{\top}\hW \bX+\bH^{\top} \bPsi \bH\right)^{-1} \bF_{d} +\bb_{2} \bb^{\top}_{2}  
\end{equation} 
where $\bb_{2} =(d-1)\left(\bX^{\top}\hW \bX+\bI\right)^{-1} \bbeta $.

\noindent Now we consider the following difference
\begin{eqnarray} \label{d2} 
\bDelta _{2} &=& MSEM\left[\srle \right]-MSEM\left[\srlte\left(k,d\right)\right] \no \\ &=& \bF_{d} \left(\bX^{\top}\hW \bX+\bH^{\top} \bPsi \bH \right)^{-1} \bF_{d} -\bF_{kd} \left(\bX^{\top}\hW \bX+\bH^{\top} \bPsi \bH \right)^{-1} \bF_{kd} +\bb_{2} \bb^{\top}_{2} -\bb_{1} \bb^{\top}_{1}\no \\ 
&=& \bD_{2} +\bb_{2} \bb^{\top}_{2} -\bb_{1} \bb^{\top}_{1} 
\end{eqnarray} 
where $\bD_2=\bF_{d} \left(\bX^{\top}\hW \bX+\bH^{\top} \bPsi \bH \right)^{-1} \bF_{d} -\bF_{kd} \left(\bX^{\top}\hW \bX+\bH^{\top} \bPsi \bH \right)^{-1} \bF_{kd}$.

Since $\bF_{d} \left(\bX^{\top}\hW \bX+\bH^{\top} \bPsi \bH\right)^{-1} \bF_{d} >\bzero$ and  $\bF_{kd} \left(\bX^{\top}\hW \bX+\bH^{\top} \bPsi \bH\right)^{-1} \bF_{kd} >\bzero$, then by Lemma \ref{rao}, when $\lambda _{\max } \left\{\bF_{kd} \left(\bX^{\top}\hW \bX+\bH^{\top} \bPsi \bH \right)^{-1} \bF_{kd} \left[\bF_{d} \left(\bX^{\top}\hW \bX+\bH^{\top} \bPsi \bH\right)^{-1} \bF_{d} \right]^{-1} \right\}<1$, $\bD_{2} >\bzero$. Thus by Lemma \ref{lem3}, when $\bb^{\top}_{1} \left(\bD_{2} +\bb_{2} \bb^{\top}_{2} \right)^{-1} \bb_{1} \le 1$, $\bDelta _{2} \ge \bzero$.

\noindent Based on the computations above, we may write the following theorem:

\begin{theorem} 
When 
$$\lambda _{\max } \left\{\bF_{kd} \left(\bX^{\top}\hW \bX+\bH^{\top} \bPsi \bH \right)^{-1} \bF_{kd} \left[\bF_{d} \left(\bX^{\top}\hW \bX+\bH^{\top} \bPsi \bH\right)^{-1} \bF_{d} \right]^{-1} \right\}<1,$$
the new estimator is superior to the SRLE if and only if $\bb^{\top}_{1} \left(\bD_{2} +\bb_{2} \bb^{\top}_{2} \right)^{-1} \bb_{1} \le 1$.
\end{theorem}

\subsection{The new estimator versus LTE}

\noindent The asymptotic properties of the LTE are given by
\begin{equation} \label{e_lte} 
E\left[\lte\left(k,d\right)\right]=\bF_{kd} \bbeta =\left(\bX^{\top}\hW \bX+k\bI\right)^{-1} \left(\bX^{\top}\hW \bX-d\bI\right)\bbeta  
\end{equation} 
and
\begin{equation} \label{cov_lte} 
Cov\left[\lte \left(k,d\right)\right]=\bF_{kd} \left(\bX^{\top}\hW \bX\right)^{-1} \bF_{kd}  
\end{equation} 
Then we have 
\begin{equation} \label{mmse_lte} 
MSEM\left[\lte \left(k,d\right)\right]=\bF_{kd} \left(\bX^{\top}\hW \bX\right)^{-1} \bF_{kd} +\bb_{1} \bb^{\top}_{1}  
\end{equation} 
where $\bb_{1} =(d+k)\left(\bX^{\top}\hW \bX+k\bI\right)^{-1} \bbeta $.
Thus, the following difference is obtained

\begin{eqnarray} \label{d3}
\Delta _{3} &=& MSEM\left[\lte \right]-MSEM\left[\srlte\left(k,d\right)\right] \no \\
&=&\bF_{kd} \left(\bX^{\top}\hW \bX\right)^{-1} \bF_{kd} -\bF_{kd} \left(\bX^{\top}\hW \bX+\bH^{\top} \bPsi \bH \right)^{-1} \bF_{kd} \no \\
&=& \bF_{kd} \left[\left(\bX^{\top}\hW \bX\right)^{-1} -\left(\bX^{\top}\hW \bX+\bH^{\top} \bPsi \bH \right)^{-1} \right]\bF_{kd}.
\end{eqnarray} 
 
By \citet{rao}, the following identity holds: 
$$\left(\bX^{\top}\hW \bX+\bH^{\top} \bPsi \bH\right)^{-1} =\left(\bX^{\top}\hW \bX\right)^{-1} -\left(\bX^{\top}\hW \bX\right)^{-1} \bH^{\top}\left[H\left(\bX^{\top}\hW \bX\right)^{-1} \bH^{\top}\right]^{-1} \bH\left(\bX^{\top}\hW \bX\right)^{-1} $$
Then $$\left(\bX^{\top}\hW \bX\right)^{-1} \bH^{\top}\left(\bH\left(\bX^{\top}\hW \bX\right)^{-1} \bH\right)^{-1} \bH\left(\bX^{\top}\hW \bX\right)^{-1} =\left(\bX^{\top}\hW \bX\right)^{-1} -\left(\bX^{\top}\hW \bX+\bH^{\top} \bPsi \bH \right)^{-1} \ge \bzero$$

\noindent So $\bDelta _{3} = MSEM\left[\lte \right]-MSEM\left[\srlte \left(k,d\right)\right]\ge \bzero$.

\noindent Therefore, we may present the following theorem: 

\begin{theorem}
The new estimator is always superior to the LTE.
\end{theorem}

\noindent 

\subsection{The new estimator versus LE}

\noindent The asymptotic properties of the LE are given as:
\begin{equation} \label{e_lle} 
E\left[\lle \left(d\right)\right]=\bF_{d} \bbeta =\left(\bX^{\top}\hW \bX+\bI\right)^{-1} \left(\bX^{\top}\hW \bX+d\bI\right)\bbeta  
\end{equation} 
and
\begin{equation} \label{cov_lle} 
Cov\left[\lle\left(d\right)\right]=\bF_{d} \left(\bX^{\top}\hW \bX\right)^{-1} \bF_{d}.  
\end{equation} 
Then we have
\begin{equation} \label{mmse_lle} 
MSEM\left[\lle \left(d\right)\right]=\bF_{d} \left(\bX^{\top}\hW \bX\right)^{-1} \bF_{d} +\bb_{2} \bb^{\top}_{2}  
\end{equation} 
where $\bb_{2} =(d-1)\left(\bX^{\top}\hW \bX+\bI\right)^{-1} \bbeta $.

\noindent Now using \eqref{mmse_lle} and \eqref{mmse_srlte} we consider the following difference

\begin{eqnarray} \label{d4} 
\bDelta _{4} &=& MSEM\left[\lle\right]-MSEM\left[\srlte \left(k,d\right)\right] \no \\ &=& \bF_{d} \left(\bX^{\top}\hW \bX\right)^{-1} \bF_{d} -\bF_{kd} \left(\bX^{\top}\hW \bX+\bH^{\top} \bPsi \bH \right)^{-1} \bF_{kd} +\bb_{2} \bb^{\top}_{2} -\bb_{1} \bb^{\top}_{1}  \no \\ 
&=& \bD_{3} -\bb_{1} \bb^{\top}_{1} .
\end{eqnarray} 
where $\bD_{3} = \bF_{d} \left(\bX^{\top}\hW \bX\right)^{-1} \bF_{d} -\bF_{kd} \left(\bX^{\top}\hW \bX+\bH^{\top} \bPsi \bH \right)^{-1} \bF_{kd}$.
Since $\bF_{d} \left(\bX^{\top}\hW \bX \right)^{-1} \bF_{d} > \bzero$ and $\bF_{kd} \left(\bX^{\top}\hW \bX+\bH^{\top} \bPsi \bH \right)^{-1} \bF_{kd} >\bzero$, then by Lemma \ref{rao}, when $$\lambda _{\max } \left\{\bF_{kd} \left(\bX^{\top}\hW \bX+\bH^{\top} \bPsi \bH \right)^{-1} \bF_{kd}  \left[\bF_{d} \left(\bX^{\top}\hW \bX \right)^{-1} \bF_{d} \right]^{-1} \right\}<1,$$
$\bD_{3} >0$. Thus by Lemma \ref{lem3}, when $\bb^{\top}_{1} \left(\bD_{3} +\bb_{2} \bb^{\top}_{2} \right)^{-1} \bb_{1} \le 1$, $\bDelta _{4} \ge \bzero$.

\noindent As a result, we can write the following theorem:

\begin{theorem}
When $\lambda _{\max } \left\{\bF_{kd} \left(\bX^{\top}\hW \bX+\bH^{\top} \bPsi \bH \right)^{-1} \bF_{kd}  \left[\bF_{d} \left(\bX^{\top}\hW \bX \right)^{-1} \bF_{d} \right]^{-1} \right\}<1$, the new estimator is superior to the LE if and only if $\bb^{\top}_{1} \left(\bD_{3} +\bb_{2} \bb^{\top}_{2} \right)^{-1} \bb_{1} \le 1$.
\end{theorem}

\subsection{Selection of the parameters k and d}

\noindent We propose an algorithm to choose the biasing parameters $k$ and $d$ iteratively for use in SRLTE. Since $\bX^{\top}\hW \bX>\bzero$, using the spectral decomposition of the matrix $\bX^{\top}\hW \bX=\bQ\bLambda \bQ^{\top}$, where $\bQ$ is the matrix whose columns are the eigenvectors of $\bX^{\top}\hW \bX$ and $\bLambda = diag\left(\lambda _{1} ,...,\lambda _{p+1} \right)$ such that $\lambda_i$'s are eigenvalues of $\bX^{\top}\hW \bX$, then we have 
\begin{equation} \label{mse_srlte}
SMSE\left[\srlte\left(k,d\right)\right]=\sum _{i=1}^{p+1}\frac{\left(\lambda _{i} -d\right)^{2} b_{ii} +\left(d+k\right)^{2} \alpha _{i}^{2} }{\left(\lambda _{i} +k\right)^{2} }   
\end{equation} 
where $\left(\bX^{\top}\hW \bX+\bH^{\top} \bPsi \bH\right)^{-1} =\bQ diag\left(b_{11} ,...,b_{p+1p+1} \right)\bQ^{\top}$ and $\hat{\alpha}=\bQ^{\top}\mle $.  


To estimate $k$, we start by taking the derivative of Equation \eqref{mse_srlte} with respect to $k$ and equating the resulting function to zero then solving for each individual parameter, we obtain the following:
\begin{eqnarray}\label{k_est}
k = \frac{(\lambda_i-d)b_{ii}-d}{\alpha _i^2}.
\end{eqnarray}
Since $k$ is positive by definition, the numerator of Equation \eqref{k_est} gives us an estimate of $d$ as

\begin{eqnarray}\label{in_d}
\frac{\lambda_i b_{ii}}{b_{ii} +1} \ge d
\end{eqnarray} 
Thus, we propose the following algorithm:
\begin{enumerate}
\item
Find an initial estimate of $d$ using \eqref{in_d} by choosing the minimum of those values.
\item
Compute the individual $k$ values using \eqref{k_est} and compute their minimum as an estimate of $k$.
\end{enumerate}

\section{A simulation study}
In this section we present the details of a Monte Carlo simulation study which is conducted to evaluate the performances of the estimators MLE, LE, LTE, SRE, SRLE and SRLTE. We consider two criteria which are the simulated mean squared error (MSE) and predictive mean MSE (PMSE) to compare the performances of listed estimators. Following \citet{MG1975}, and \citet{VW2016b}, we generate the data matrix $\bX$ containing the explanatory variables such that $\rho^2$ is the degree of collinearity between any two variables as follows:

\begin{equation}
x_{ij} = (1-\rho^2)^{1/2}w_{ij} + \rho w_{i(p+1)}
\end{equation}
where $w_{ij}$ are obtained from the standard normal distribution. The data matrix is centered and standardized before computations.

We consider the following setting:
\begin{itemize}
\item
The sample size is taken as $50,100$ and $200$.
\item
The number of explanatory variables is $p=4$.
\item
The parameter values as chosen such that $\bbeta^{\top}\bbeta=1$. 
\item
The degree of correlation $\rho$ changes as $0.9, 0.99$ and $0.999$.
\item
The dependent variable is generated using
$\pi_{i} = \frac{e^{x_{i}} \bbeta}{1+e^{x_{i}} \bbeta } $

\item
The restrictions are chosen as follows:
Following \citet{VW2016b}, we choose 
$\bH = 
\left[ {\begin{array}{*{20}c}
   1 & -1 &0&0 \\
   0 & 1 & -1&0  \\
   0&0&1&-1\\
 \end{array} } \right]
$ 
and $\bh=\bH\bbeta +\bv$ where $\bv \sim N(0, \bPsi)$ such that $\bPsi = \bI_3$.

\item
We choose the biasing parameter of LE and SRLE following \citet{Manson2012}. For LTE, we follow \citet{Asar2016} to choose $k$ and $d$. For SRLTE, we use our proposed algorithm to estimate the parameters.

\end{itemize}
\begin{table}[ht]
\caption{The simulated MSE values}
\centering
\begin{tabular}{rrrrrrr}
  \hline
n & MLE & LE & LTE & SRMLE & SRLE & SRLTE \\ 
      \hline
  \multicolumn{7}{c}{$\rho=0.9$}     \\
  \hline
50 & 1.85652 & 1.36613 & 1.12453 & 0.51381 & 0.66333 & 0.58042 \\ 
  100 & 1.78156 & 1.34392 & 1.10278 & 0.58567 & 0.74037 & 0.67935 \\ 
  200 & 1.75873 & 1.50912 & 1.16144 & 0.86342 & 0.89278 & 0.91667 \\ 
      \hline
  \multicolumn{7}{c}{$\rho=0.99$}     \\
  \hline
  50 & 2.60743 & 1.03528 & 1.20035 & 0.72942 & 0.93741 & 0.71122 \\ 
  100 & 2.18982 & 1.05415 & 1.13695 & 0.67586 & 0.90969 & 0.65671 \\ 
  200 & 1.89717 & 1.11810 & 1.08619 & 0.60522 & 0.82612 & 0.58546 \\ 
      \hline
  \multicolumn{7}{c}{$\rho=0.999$}     \\
  \hline
  50 & 17.24458 & 1.00220 & 3.61286 & 0.82232 & 0.99678 & 0.82139 \\ 
  100 & 5.79611 & 1.00791 & 1.78444 & 0.81379 & 0.98176 & 0.80758 \\ 
  200 & 3.41868 & 1.01677 & 1.36106 & 0.77453 & 0.96657 & 0.76399 \\ 
   \hline
\end{tabular}
\end{table}
\begin{table}[ht]
\caption{The simulated PMSE values}
\centering
\begin{tabular}{rrrrrrr}
  \hline
n & MLE & LE & LTE & SRMLE & SRLE & SRLTE \\ 
        \hline
  \multicolumn{7}{c}{$\rho=0.9$}     \\
  \hline
50 & 2.66301 & 2.42390 & 2.49373 & 2.35958 & 2.33175 & 2.34593 \\ 
  100 & 4.80013 & 4.58821 & 4.65525 & 4.56813 & 4.54326 & 4.55901 \\ 
  200 & 9.63524 & 9.22454 & 9.45762 & 9.13080 & 9.11585 & 9.12089 \\
        \hline
  \multicolumn{7}{c}{$\rho=0.99$}     \\
  \hline 
  50 & 2.32046 & 2.25402 & 2.20685 & 2.22910 & 2.19648 & 2.22816 \\ 
  100 & 4.49735 & 4.42789 & 4.38679 & 4.41898 & 4.37117 & 4.41659 \\ 
  200 & 8.87562 & 8.78514 & 8.76574 & 8.80102 & 8.73969 & 8.79209 \\ 
        \hline
  \multicolumn{7}{c}{$\rho=0.999$}     \\
  \hline
  50 & 2.30213 & 2.25291 & 2.18845 & 2.19244 & 2.18757 & 2.19244 \\ 
  100 & 4.47697 & 4.41158 & 4.36401 & 4.36879 & 4.36197 & 4.36873 \\ 
  200 & 8.81881 & 8.76106 & 8.70836 & 8.72060 & 8.70271 & 8.72036 \\ 
   \hline
\end{tabular}
\end{table}
The simulation is repeated $5000$ times. All computations are performed using the R program (\citet{R2016}). The simulated MSE and PMSE values of estimators are computed respectively as follows:
 $$
\mbox{MSE}(\tilde{\beta})=\frac{\sum_{r=1}^{5000}(\tilde{\bbeta}_r-\beta)^{\top}(\tilde{\bbeta}_r-\beta)}{5000}
     $$
     and 
   $$
\mbox{PMSE}(\tilde{\beta})=\frac{\sum_{r=1}^{5000}(\tilde{\bpi}_r-\bpi)^{\top}(\tilde{\bpi}_r-\bpi)}{5000}
     $$   
where $\tilde{\beta}_r$  is any estimator considered in the study and $\tilde{\bpi}_r$ is the predicted probabilities in
the $r^{th}$ repetition.

The results of the simulation are presented in Tables 1-4. According to the results,  we observe that SRLTE has the best performance in the sense of MSE for most of the situations and MLE has the worst performance in the sense of both MSE and PMSE. When $\rho=0.9$, SRE has the best performance, however, SRLTE becomes the second best estimator.
It was expected that SRLE would have a better performance than SRE, but it is not. This may be the result of the selection of the biasing parameter.

Although, MSE of LTE is lower than MSE of LE and MLE when the correlation is low, LE has a better performance than LTE when the correlation is high.

It is also seen from tables that increasing the degree of correlation makes an increase in the MSE values of the estimators except for LTE and LE. However, if we consider PMSE values, this situation is vice versa, namely, if the degree of correlation increases, PMSE values decrease. According to Table 2, it is observed that SRLE has the best predictive performance. SRLTE is the second best estimator in the sense of PMSE. 

Moreover, if we consider MSE as a function of the sample size, we conclude that all the estimators have asymptotic properties for most of the cases, in other words, if the sample size increases, MSE values decrease. On the other hand the PMSE values increase for the same situation.


\section{Conclusion}
In this paper, we propose a new stochastic restricted estimator SRLTE and study its performance both theoretically and numerically. We obtain some conditions such that SRLTE is superior to SRE, SRLE, LTE and LE using mean square error matrices of estimators. We also propose an algorithm to choose the biasing parameters of SRLTE. Moreover, we conduct a Monte Carlo simulation experiment to compare the performances of estimators numerically. We used both MSE and predictive MSE as comparison criteria. According to the results of the simulation, SRLTE performs better in most of the situations in the sense of MSE.

\section*{Acknowledgement}
This research is supported by...

\end{document}